\documentclass[aps,prl,twocolumn,superscriptaddress,showpacs,amsmath,floatfix,citeautoscript]{revtex4-2}

\usepackage[utf8]{inputenc}
\usepackage{graphicx}
\usepackage{mathtools}
\usepackage{xcolor}
\usepackage{amsmath}
\usepackage{amssymb}
\usepackage{xfrac}
\usepackage{soul}

\begin{document}
	
	\title{Generalized Neumann’s Principle as a Unified Framework for Fractional Quantum and Conventional Ferroelectricity}
	
	\author{Hongsheng Pang}
	\affiliation{CAS Key Laboratory of Quantum Information, University of Science and Technology of China, Hefei 230026, People's Republic of China}
	
	\author{Lixin He}
	\email{helx@ustc.edu.cn}
	\affiliation{CAS Key Laboratory of Quantum Information, University of Science and Technology of China, Hefei 230026, People's Republic of China}
	\affiliation{Institute of Artificial Intelligence, Hefei Comprehensive National Science Center, Hefei, 230088, People's Republic of China}
	\affiliation{Hefei National Laboratory, University of Science and Technology of China, Hefei 230088, China}
	
	\date{\today}
	
	\begin{abstract}	
		Monolayer In$_2$Se$3$ exhibits unexpected in-plane polarization, despite having $C_{3v}$ symmetry, a feature that was traditionally considered forbidden by symmetry. To explain this remarkable behavior, Ji et al. proposed the concept of fractional quantum ferroelectricity (FQFE), in which polarization occurs in fractional multiples of a quantum, and argued that this phenomenon violates {\it conventional} Neumann's principle.
In this Letter, we introduce a generalized form of Neumann’s principle and demonstrate that both FQFE and conventional ferroelectricity can be consistently described within this unified theoretical framework.
		We propose a method, based on the generalized Neumann's principle, for the systematic identification of FQFE materials.
This approach is straightforward to apply and offers a clear conceptual understanding and deep physical insight for FQFE.   Using this method, we determine all symmetry-allowed FQFE cases across the 32 crystallographic point groups.
		Since practical applications rely on the ability to control polarization, we further show that FQFE can be effectively switched via coupling with conventional polarization. Using HfZnN$_2$ as an illustrative example, we reveal the underlying mechanism of this coupling and outline a strategy to identify other materials with similar switching behavior.
	\end{abstract}

	\date{\today}	
	\maketitle
	
	
	Recent investigations into two-dimensional materials have revealed intriguing phenomena that challenge established theoretical frameworks. Among these materials, the monolayer In$_2$Se$_3$ has attracted significant attention due to its unconventional electric polarization behavior. 	
	Despite possessing $C_{3v}$ symmetry, which typically forbids in-plane polarization, In$_2$Se$_3$ exhibits an unexpectedly large in-plane polarization, as observed in experiments~\cite{InSe-FE,In2Se3-exp2,In2Se3-exp3,In2Se3-exp4}. This surprising behavior contradicts conventional symmetry-based expectations and has sparked extensive discussions regarding its underlying physical origin, with many attributing the in-plane polarization to  lattice asymmetry~\cite{ding2017prediction,InSe-FE2,InSe-FE}.
	
To explain this phenomenon, Ji et al.~\cite{FQFE,PRL2025} proposed the concept of fractional quantum ferroelectricity (FQFE), where polarization occurs in fractional multiples of a quantum. They also introduced a systematic methodology for identifying materials exhibiting FQFE and reported a large number of such materials~\cite{FQFE,PRL2025}.
Furthermore, they suggested that FQFE violates (conventional) Neumann's principle~\cite{neumann1885vorlesungen}, a widely accepted criterion for identifying ferroelectric materials. This reveals a conceptual gap between conventional ferroelectrics and FQFE, underscoring the need to extend existing theories into a more general framework capable of simultaneously describing both types of ferroelectrics.
	
In this Letter, we present a unified theoretical framework for understanding both FQFE materials and conventional ferroelectrics, grounded in a generalized form of Neumann's principle, along with a corresponding method for identifying such materials.
Our approach is straightforward to apply, conceptually clear, and provides deep insights into the underlying physics. Using this approach, we have identified several new FQFE materials. One such material, HfZnN$_2$, exhibits out-of-plane conventional polarization and in-plane FQFE. We demonstrate that coupling to an external perpendicular electric field facilitates the switching of the in-plane polarization in this material, offering an effective means of controlling FQFE.

\begin{figure}[tbp]
		\centering
		\includegraphics[width=1\linewidth]{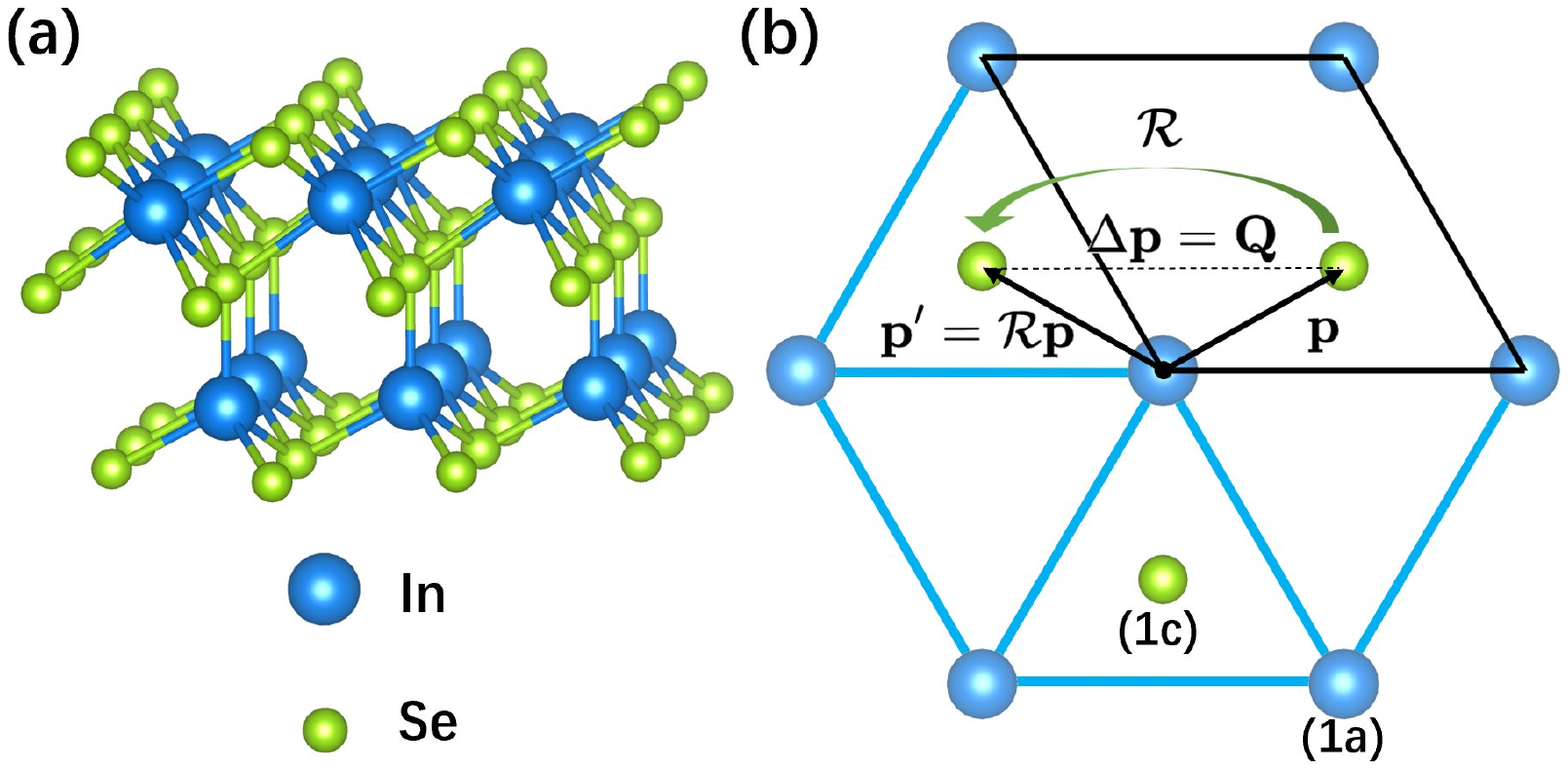}
		\caption{(a) The structure of ZB $\alpha$-In$_2$Se$_3$. (b) A schematic illustration of the generalized Neumann's principle, using In$_2$Se$_3$ as an example. Only the In atom at the 1a Wyckoff site and the Se atom at the 1c site are shown from the top view of the lattice.
			Solid black lines represent lattice boundaries, and black arrows show in-plane FQFE for $\mathbf{p}$ and $\mathbf{p}'$. A symmetry operation $\mathcal{R}$ transforms $\mathbf{p}$ to $\mathbf{p}'$, where the dotted black line indicates a difference of exactly one quantum.
		}
		\label{symmetry}
	\end{figure}

		
In addition to its other remarkable properties, the zinc-blende (ZB) phase of In$_2$Se$_3$ has been reported to exhibit ferroelectricity~\cite{FQFE, layer-dependent-FE, ding2017prediction}. The ZB structure of In$_2$Se$_3$, shown in Fig.~\ref{symmetry}(a), belongs to the $C_{3v}$ point group, which lacks inversion symmetry and therefore permits out-of-plane electric polarization.
	However, experiments~\cite{In2Se3-exp2, In2Se3-exp3, In2Se3-exp4, InSe-FE, InSe-FE2} have also observed a significant in-plane polarization, which is puzzling because it contradicts conventional expectations: the $C_{3v}$ symmetry should, in principle, forbid any in-plane component of the polarization.
	Ji et al. propose that the material can exhibit an in-plane polarization of $(2/3, 1/3)$ of a quantum, which they refer to as FQFE. Furthermore, they proposed a systematic method for identifying materials exhibiting FQFE and  reported numerous materials with this property\cite{FQFE,PRL2025}.
	
The existence of nonzero fractionally quantized nominal polarization in high-symmetry crystals was previously discussed by Vanderbilt et al.~\cite{vanderbilt2018berry, rabe2007physics}.
For example, in the GaAs structure, the polarization takes the value $(1/4,\,1/4,\,1/4)$ in units of a quantum. However, since these fractional quantum polarizations were generally believed to be irreversible, they did not attract significant attention at the time.
At first glance, this result seems to contradict physical intuition. The apparent paradox arises from the fact that electric polarization is intrinsically multivalued modulo a quantum~\cite{vanderbilt1993electric}.

The symmetry of a crystal structure is characterized by both its point group and translational symmetries~\cite{kittel2018introduction}. As a consequence, electric polarization is only well-defined up to integer multiples of a quantum~\cite{vanderbilt2018berry,rabe2007physics}.
Therefore, under a point group operation, the polarization may change by an integer multiple of a quantum, i.e.,
	\begin{equation}
		\mathcal{R}\mathbf{p} = \mathbf{p} + \mathbf{Q} ,
		\label{eq:neum}
	\end{equation}
	where $\mathcal{R}$ represents the symmetry operation of the crystal, $\mathbf{p}$ is the polarization vector, and $\mathbf{Q}$ is the additional quanta arising from the translation symmetry.
	This is analogous to quasi-momentum conservation in crystals, where the change in momentum $\mathbf{k}$ is defined only modulo a reciprocal lattice vector $\mathbf{G}$, i.e., $\mathbf{k} \to \mathbf{k} + \mathbf{G}$, where $\mathbf{k}$ is the quasi-momentum of a particle and $\mathbf{G}$ is a reciprocal lattice vector~\cite{kittel2018introduction}.
	Similarly, in the case of polarization, the shift by $\mathbf{Q}$ arises from the translational symmetry of the crystal.
	
	We show that FQFE can be naturally explained within the framework of this generalized Neumann’s principle, as given in Eq.~(\ref{eq:neum}).
	It is straightforward to verify that the in-plane polarization of In$_2$Se$_3$ satisfies Eq.~(\ref{eq:neum}). Consider, for instance, an in-plane polarization of $\mathbf{p}_\parallel = (2/3,\,1/3)$.
	The three-fold rotation matrix in the $ab$ plane \cite{bilbao} is:
	$\mathcal{R} =
	\begin{pmatrix}
		0 & -1 \\
		1 & -1
	\end{pmatrix}$.
	Applying the rotation matrix $\mathcal{R}$ to the polarization vector $\mathbf{p}_\parallel$, we obtain
	$\mathcal{R}  \mathbf{p}_\parallel = (-1/3,\,1/3) = \mathbf{p}_\parallel + \mathbf{Q}$,
	where $\mathbf{Q} = (-1,\,0)$ is a polarization quantum arising from the translational symmetry. This confirms that the polarization satisfies Eq.~(\ref{eq:neum}).
	This process is illustrated in Fig.~\ref{symmetry}(b), where a $120^\circ$ rotation of the polarization results in a change of exactly one quantum.
	One can readily verify that this $\mathbf{p}_\parallel$ satisfies Eq.~(\ref{eq:neum}) under all symmetry operations of the $C_{3v}$ point group, confirming that $\mathbf{p}_\parallel = (2/3,\,1/3)$ is a symmetry-allowed in-plane polarization in accordance with Neumann's principle. Similarly, $\mathbf{p}'_\parallel = (1/3,\,2/3)$ is also permitted.
	
In In$_2$Se$_3$, the in-plane polarization is coupled to the out-of-plane polarization. When the out-of-plane component $\mathbf{p}_\perp$ is reversed along the $z$-axis to $-\mathbf{p}_\perp$, the in-plane polarization in the $xy$ plane changes from $\mathbf{p}_\parallel = (2/3,\,1/3)$ to $\mathbf{p}'_\parallel = (1/3,\,2/3)$, resulting in a polarization difference of $\Delta \mathbf{p}_\parallel = \mathbf{p}'_\parallel - \mathbf{p}_\parallel = (-1/3,\,1/3)$.
	
	
	We now show that the generalized Neumann’s principle can be used to identify both FQFE and conventional FE states. We reformulate Eq.~(\ref{eq:neum}) as follows:
	\begin{equation}
		(\mathcal{R} - \mathcal{I}) \mathbf{p} = \mathbf{Q},
		\label{eq:linalgeq}
	\end{equation}
	where this equation implies that, under any symmetry operation $\mathcal{R}$ of the symmetry group, the polarization vector $\mathbf{p}$ may change by an integer quantum $\mathbf{Q}$. Solving this linear equation allows us to determine all symmetry-allowed polarization vectors $\mathbf{p}$ for a given symmetry group.
	
	We present a systematic approach to identify all possible FQFE solutions for a given point group. First, we select a symmetry operation $\mathcal{R}$ from the point group and solve the linear equation, Eq.~(\ref{eq:linalgeq}), to obtain a candidate set of polarization vectors $\{\mathbf{p}_i\}$. Each candidate polarization vector is then tested against all other symmetry operations in the point group to verify whether it satisfies Eq.~(\ref{eq:linalgeq}). Only those vectors that satisfy the condition for all symmetry operations are considered valid FQFE solutions, in accordance with the generalized Neumann’s principle.
	
We formulate the equation in the direct crystal coordinate system. Since the polarization vector $\mathbf{p}$ is typically small, we restrict the components of $\mathbf{Q}$ to the values $-1$, $0$, and $1$.
The case of $\mathbf{Q} = 0$ corresponds to the {\it conventional} Neumann's principle, which has traditionally been applied to determine whether the symmetry of a crystal structure permits ferroelectricity.
Additional constraints may apply to the permissible values of $\mathbf{Q}$.
For example, in the case of the $C_1$ point group, the system possesses no symmetry other than the identity, so $\mathcal{R} - \mathcal{I} = \mathbf{0}$. In this case, for Eq.~(\ref{eq:linalgeq}) to admit a solution, $\mathbf{Q}$ must be zero, and $\mathbf{p}$ can take arbitrary values, corresponding to conventional polarization.
More generally, in polar point groups, where the polarization remains invariant under all symmetry operations
of the group, the matrix $\mathcal{R} - \mathcal{I}$ becomes rank-deficient. As a result, the corresponding components of $\mathbf{Q}$ must be zero, reflecting the existence of unconstrained (i.e., continuous) polarization components along the polar directions.
	
	As an illustrative example, we again consider In$_2$Se$_3$, which exhibits $C_{3v}$ symmetry. When applying the three-fold rotation about the $c$-axis, we obtain:
	\begin{equation}
		\mathcal{R} - \mathcal{I} = \begin{pmatrix}
			0 & -1 & 0 \\
			1 & -1 & 0 \\
			0 & 0 & 1
		\end{pmatrix} -
		\begin{pmatrix}
			1 & 0 & 0 \\
			0 & 1 & 0 \\
			0 & 0 & 1
		\end{pmatrix} =
		\begin{pmatrix}
			-1 & -1 & 0 \\
			1 & -2 & 0 \\
			0 & 0 & 0
		\end{pmatrix}
	\end{equation}
	It is straightforward to verify that the rank of the matrix $(\mathcal{R} - \mathcal{I})$ is two.
	Consequently, to satisfy Eq.~(\ref{eq:linalgeq}), $Q_c$ must be zero, leaving the out-of-plane polarization $p_c$ unconstrained.
	By excluding the null space, we solve the linear equation. The special solutions for $\mathbf{Q} = (-1,\,0,\,0)$ are given by $\mathbf{p} = (2/3,\,1/3,\,p_c)$, where $p_c$ can take any arbitrary value. The in-plane polarization $\mathbf{p}_\parallel = (2/3,\,1/3)$ corresponds to the FQFE discussed in the previous text. Similarly, applying $\mathbf{Q} = (1,\,0,\,0)$ yields $\mathbf{p}' = (1/3,\,2/3,\,p_c)$, where the in-plane component $\mathbf{p}'_\parallel = (1/3,\,2/3)$. These are the two inequivalent FQFE solutions allowed by $C_{3v}$ symmetry.
	One can readily verify that the solutions for other $\mathbf{Q}$ values are equivalent to these two, differing only by integer quanta. Furthermore, both solutions satisfy Eq.~(\ref{eq:linalgeq}) under all symmetry operations of the $C_{3v}$ group.
	
	For non-polar groups, the matrix $(\mathcal{R} - \mathcal{I})$ is of full rank, therefore, one can directly solve the linear equations, and the resulting $\mathbf{p}$ values correspond to fractional quantum.
	We emphasize that this method identifies only the symmetry-allowed polarizations. To confirm the existence of FQFE in a specific material, one must still perform first-principles calculations, as the trivial solution $\mathbf{p} = (0, 0, 0)$ always satisfies Eq.~(\ref{eq:linalgeq}).
	
	We have applied this scheme to explore potential FQFE states across all 32 crystallographic point groups. A comprehensive summary of the FQFE states is presented in Table~S1 of  Supplemental Materials (SM) \cite{SM} for all standard point group operations~\cite{bilbao}. In this table, we list all inequivalent polarization values, except the trivial zero solution. In the ten polar point groups, the polarization is allowed to take arbitrary values along the polar axis, while in non-polar groups, FQFE polarizations assume discrete values due to the symmetry constraints imposed by Eq.~(\ref{eq:linalgeq}). Moreover, any linear combination of the allowed polarizations with integer coefficients also satisfies the generalized Neumann's principle.
	
	We calculate the polarization of 20 materials of different symmetries, and the results are listed in Table S3 of SM~\cite{SM}.
	It should be noted that the nominal values of FQFE may vary depending on the choice of unit cell, as this can alter the symmetry operation matrices. For example, the conventional unit cell of AgBr has a face-centered cubic (fcc) structure with $T_d$ symmetry, corresponding to the standard point group operations. As listed in Table~S1, the polarization can take values of $(1/2,\,1/2,\,1/2)$ or zero (i.e., integer) quanta. First-principles calculations confirm that the polarization in the conventional cell corresponds to zero quanta.
However, if the primitive unit cell is used, which also belongs to the $T_d$ point group but involves a different set of symmetry operations (see Sec. B in SM~\cite{SM}), the resulting FQFE is $(1/4,\,1/4,\,1/4)$ quanta. While this may initially appear surprising, it is consistent with the fact that the definition of a polarization quantum depends on the choice of unit cell, specifically given by $\mathbf{Q} = \mathbf{R}/\Omega$, where $\mathbf{R}$ is a lattice vector and $\Omega$ is the unit cell volume. When this scaling factor is taken into account, the actual FQFE values obtained from the two unit cells are found to be equivalent (see Sec. B in SM~\cite{SM}).
	
For some point groups, there exist inequivalent FQFE solutions, and first-principles calculations are necessary to determine which solution is realized in a given structure. As previously discussed,  even in materials whose point group symmetry permits FQFE, the actual polarization may still vanish. For example, SrAlSiH possesses $C_{3v}$ symmetry, which allows an in-plane polarization of $(1/3,\,2/3)$ quanta~\cite{PRL2025}. However, ab initio calculations reveal that, in this compound, the ionic contribution is $(1/3,\,2/3)$ quanta, while the electronic contribution is $ (-1/3,\,-2/3)$ quanta. These contributions cancel each other out, resulting in a net zero in-plane polarization (see Table S3 in  SM~\cite{SM}).
	
When a symmetry operation beyond the native point group of a crystal is applied, the allowed polarization solution may transform accordingly.
The multiple inequivalent solutions listed in Table~S1 indicate that such transformations can map the polarization to distinct allowed states, leading to a rich and intriguing diversity of possible outcomes.
If the point group allows only a single nontrivial FQFE solution, then under an external symmetry operation, the FQFE may either shift by an integer quantum or remain unchanged.
	
However, if the point group of the material permits multiple inequivalent FQFE solutions, two scenarios may arise. In one case, upon applying a symmetry operation beyond the point group of the material, the FQFE remains within the original equivalence class, in which case the change in polarization is either zero or an integer multiple of a quantum. For example, KNiIO$_6$ belongs to the $D_3$ point group, which allows three inequivalent FQFE solutions. Ab initio calculations show that KNiO$_6$ has a polarization of $(0,\,0,\,1/2)$ quanta. Under an inversion operation, the polarization changes by one quantum to $(0,\,0,\,-1/2)$, while under a three-fold in-plane rotation, it remains unchanged. Nonetheless, in both cases, the polarization is invariant modulo a quantum.
	
In the other case, the system transitions to an inequivalent FQFE state, where the change in polarization corresponds to a fractional quantum. Exemplary materials for this case include In$_2$Se$_3$ and AgAlS$_2$, which correspond to Type-I FQFE as described in Ref.~\cite{PRL2025}.
	
For ferroelectric materials, the feasibility of practical applications critically depends on the ability to control and switch their polarization. Recent study~\cite{PRL2025}, have demonstrated that FQFE can couple directly to an external electric field.
Alternatively, polarization switching in In$_2$Se$_3$~\cite{InSe-FE} reveals that FQFE can also be controlled via coupling to conventional ferroelectric polarization. These findings demonstrate that FQFE can be tuned through multiple mechanisms, offering greater flexibility for applications and material design.
This motivates the search for materials where FQFE is switchable via coupling to conventional ferroelectricity.

To identify suitable materials, the system must exhibit non-zero FQFE and belong to one of the ten polar point groups. This criterion narrows the candidates to $C_2$, $C_s$, $C_{2v}$, $C_4$, $C_{4v}$, $C_3$, and $C_{3v}$.
We explore materials in the $C_3$ and $C_{3v}$ point groups from the Materials Project database~\cite{material-project} to identify potential FQFE candidates using our proposed scheme. We have verified the materials identified in Ref.~\cite{PRL2025}, including AgBiP$_2$Se$_6$, AgAlS$_2$, HAlSiSr, and others. However, some materials identified as FQFE actually show zero polarization upon first-principles calculations. Examples can be found in Table S3 of  SM~\cite{SM}.
Additionally, we identify several new materials, including HfZnN$_2$, SrZnN$_2$, YZnCuP$_2$, BiTeI, Nb$_3$SBr$_7$ and Ag$_6$Mo$_2$ClO$_7$F$_3$ that exhibit both in-plane FQFE and out-of-plane traditional polarization. First-principles calculations have been performed to confirm these results as presented in Table S3 and S4 of SM~\cite{SM}.
	
We use HfZnN$_2$  as an example to investigate the detailed mechanism of coupling between conventional polarization and FQFE using first-principles calculations. The calculations are performed within the framework of density functional theory (DFT), as implemented in the Atomic Orbital Based Ab-initio Computation at UStc (ABACUS) package~\cite{ABACUS1,ABACUS2}. The Perdew–Burke–Ernzerhof (PBE) exchange-correlation functional~\cite{GGA} is employed.
Optimized norm-conserving Vanderbilt (ONCV) pseudopotentials from the SG15 set~\cite{SG15,ONCV} are used, with the Hf 5$d^{2}$6$s^{2}$, Zn 3$d^{10}$4$s^{2}$, and N 2$s^{2}$2$p^{3}$ electrons treated as valence electrons~\cite{orbital,Linpz2023}. Numerical atomic orbital (NAO) basis sets of 4s2p2d2f1g for Hf, 4s2p2d1f for Zn, and 2s2p1d for N are employed~\cite{orbital}.
For self-consistent calculations, an $8 \times 8 \times 8$ mesh is adopted. The energy cutoff for the wavefunctions is set to 100~Ry.

	\begin{figure}[tbp]
		\centering
		\includegraphics[width=0.8\linewidth]{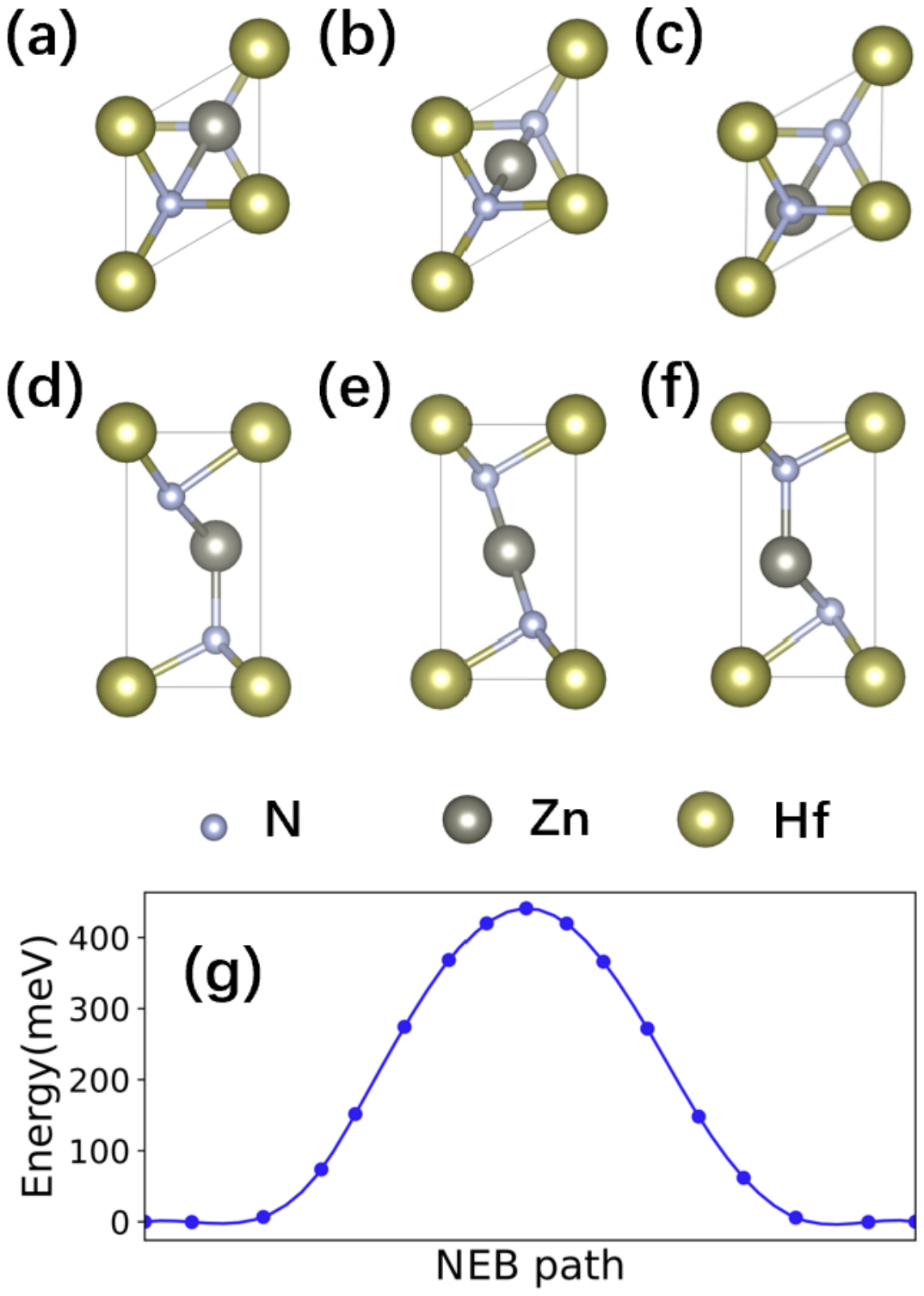}
		\caption{The NEB path of polarization reversal in HfZnN$_2$. (a)-(f) are schematic illustrations of the structural changes along the NEB path, where (a) and (d) represent the top and side views of the initial state, (b) and (e) show the intermediate state, and (c) and (f) correspond to the final state. (g) shows the energy barrier along the NEB path.}
		\label{HfZnN}
	\end{figure}

HfZnN$_2$ has the P3m1 space group with $C_{3v}$ point group symmetry. This symmetry permits a conventional polarization along the $c$-axis, as well as in-plane FQFE states of $(1/3,\,2/3)$ or $(2/3,\,1/3)$ quanta. The Hf atom occupies the high-symmetry 1a Wyckoff position~\cite{SM}.
The asymmetric atomic positions include the Zn atom at the 1b Wyckoff site and two N atoms at the 1b and 1c sites
as listed in Table S5 of SM~\cite{SM}. The chemical environments of the two N atoms are inequivalent, with the Hf--N bond length being 2.30~\AA\ for the upper N atom and 2.11~\AA\ for the lower one, giving rise to the out-of-plane conventional polarization.

Figure~\ref{HfZnN} shows the two polar states of HfZnN$_2$ and the optimized switching path, calculated using the nudged elastic band (NEB) method~\cite{neb1,neb2}, where (a),(d) represent one polar state, (c),(f) depict the second, and (b),(e) show the intermediate states.
The energy barrier for ferroelectric switching is approximately 400~meV, which is relatively low. Throughout the switching path, the Hf atom remains fixed at its high-symmetry Wyckoff position, whereas the Zn atom undergoes a displacement from the upper-right to the lower-left region of the unit cell.
In the in-plane direction, the Zn atom moves from $(1/3,\,2/3)$ (1b site) to $(2/3,\,1/3)$ (1c site). Both N atoms shift upward along the $c$-axis, resulting in a shortening of the upper Hf--N bonds and a lengthening of the lower Hf--N bonds. During the switching process, the N atoms undergo slight in-plane displacements from their initial 1b and 1c Wyckoff positions but return to these positions at the end of the switching path.

\begin{figure}[tbp]
		\centering
		\includegraphics[width=0.8\linewidth]{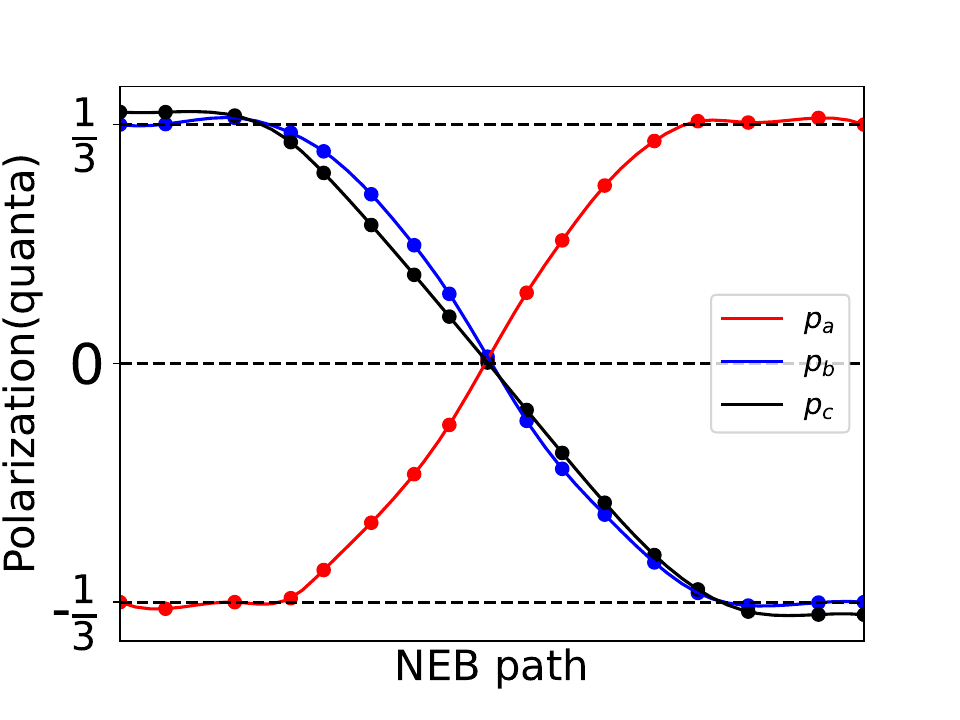}
		\caption{Polarization switching along the NEB path. The red, blue, and black lines represent the polarization, in units of a quantum, along the $a$, $b$, and $c$ axes, respectively.}
		\label{HfZnN-polar}
	\end{figure}

We compute the electric polarization along the switching path using the Berry phase method~\cite{vanderbilt2018berry}, and the results are shown in Fig.~\ref{HfZnN-polar}. The conventional polarization $p_c$ changes from 0.351 to $-0.351$ in units of a quantum, resulting in a total change of $\Delta p_c = 0.702$ quanta, which corresponds to 1.236~C/m$^2$, comparable to that in typical perovskite ferroelectric materials such as PbTiO$_3$~\cite{PbTiO-1,PbTiO-2,PbTiO-3}.

As the Zn atom moves along the $c$ axis, its motion in the $ab$ plane is coupled with this displacement, causing it to deviate from the high-symmetry Wyckoff positions. This coupling reduces the symmetry of the structure to the $C_1$ point group, allowing for the emergence of conventional polarization in the $ab$ plane. As shown in Fig.~\ref{HfZnN-polar}, along the switching path, the in-plane polarization changes continuously from $(2/3,\,1/3)$ to $(1/3,\,2/3)$ quanta, where a quantum equals 1.081 C/m$^2$ \cite{note1}.
	This symmetry reduction is crucial for polarization switching, as it allows a continuous transition between (fractionally) quantized polarization states. Without such a symmetry-lowering pathway, the polarization would have to change discontinuously between quantized values, which is physically unlikely.
	
During the switching process, the system attains a high-symmetry intermediate structure corresponding to the space group C2/m, with point group $C_{2h}$. In this phase, Hf occupies the 2a Wyckoff site, Zn the 2d site, and N the 4i site. The polarization in this intermediate state is exactly zero.

To switch the in-plane polarization, an electric field can be applied along the $c$-direction, providing an effective and flexible means of control that makes these materials more attractive for device applications. This mechanism lays the foundation for the future use of FQFE materials in advanced devices, including highly integrated ultra-thin ferroelectric memory units~\cite{InSe-FE} and ferroelectric random-access memory (FeRAM)~\cite{FeRAM}.

	
To summarize, when analyzing ferroelectric behavior in crystalline materials, it is essential  to account for  translational symmetry, which leads naturally to a generalized form of Neumann’s principle. We demonstrate that both FQFE and conventional ferroelectricity can be consistently described within a unified theoretical framework based on this principle. By applying the generalized Neumann’s principle, symmetry-allowed FQFE and conventional polarization states can be systematically identified from the point group symmetry of a given system. Furthermore, we show that coupling FQFE with conventional ferroelectricity provides an effective mechanism for polarization switching, which paves the way for its future device applications.

	
We thank D. Vanderbilt for valuable discussion. This work was supported by the Strategic Priority Research Program of Chinese Academy of Sciences (Grant Number XDB0500201), the National Natural Science Foundation of China (Grant Number 12134012),
and the Innovation Program for Quantum Science and Technology Grant Number 2021ZD0301200.
The numerical calculations were performed on the USTC HPC facilities.


\begin{thebibliography}{35}
\bibitem [{not()}]{note1}%
  \BibitemOpen
  \href@noop {} {}\bibinfo {note} {In Sec.~E of the SM \cite{SM}, we present an
  alternative polarization reversal path, where the polarization change can
  reach $4/3$ quanta. However, the energy barrier along this path is
  significantly higher than that of the current pathway, rendering it
  practically inaccessible.}\BibitemShut {Stop}%
\end{thebibliography}
%

\end{document}